\documentclass[aps,prb,reprint,
groupedaddress,
superscriptaddress,twocolumn,showpacs,
amsmath,amssymb]{revtex4-1}
\usepackage{graphicx}
\usepackage{dcolumn}
\usepackage{bm}
\usepackage{color}
\usepackage{graphicx}
\usepackage{upgreek}
\begin{document}
\title{Anisotropic magnetization relaxation in ferromagnetic multilayers with variable interlayer exchange coupling}
\author{A. F. Kravets}
\email{anatolii@kth.se}
\thanks{Correspondence author}
\affiliation{Institute of Magnetism, National Academy of Sciences of Ukraine, 36 b Vernadsky Ave., 03680 Kyiv, Ukraine}
\affiliation{Nanostructure Physics, Royal Institute of Technology, 10691 Stockholm, Sweden}
\author{D. M. Polishchuk}
\affiliation{Institute of Magnetism, National Academy of Sciences of Ukraine, 36 b Vernadsky Ave., 03680 Kyiv, Ukraine}
\affiliation{Nanostructure Physics, Royal Institute of Technology, 10691 Stockholm, Sweden}
\author{Yu. I. Dzhezherya}
\author{A. I. Tovstolytkin}
\author{V. O. Golub}
\affiliation{Institute of Magnetism, National Academy of Sciences of Ukraine, 36 b Vernadsky Ave., 03680 Kyiv, Ukraine}
\author{V. Korenivski}
\affiliation{Nanostructure Physics, Royal Institute of Technology, 10691 Stockholm, Sweden}
\date{\today}

\begin{abstract}
The FMR linewidth and its anisotropy in F$_1$/f/F$_2$/AF multilayers, where spacer f has a low Curie point compared to the strongly ferromagnetic F$_1$ and F$_2$, is investigated. The role of the interlayer exchange coupling in magnetization relaxation is determined experimentally by varying the thickness of the spacer. It is shown that stronger interlayer coupling via thinner spacers enhances the microwave energy exchange between the outer ferromagnetic layers, with the magnetization of F$_2$ exchange-dragged by the resonance precession in F$_1$. A weaker mirror effect is also observed: the magnetization of F$_1$ can be exchange-dragged by the precession in F$_2$, which leads to anti-damping and narrower FMR linewidths. A theory is developed to model the measured data, which allows separating various contributions to the magnetic relaxation in the system. Key physical parameters, such as the interlayer coupling constant, in-plane anisotropy of the FMR linewidth, dispersion of the magnetic anisotropy fields are quantified. These results should be useful for designing high-speed magnetic nanodevices based on thermally-assisted switching.
\end{abstract}
%
\maketitle

\section{Introduction}
Recent years have shown a growing research interest in advanced magnetic multilayers due to their high potential for applications in magnetic random access memory and microwave devices.\cite{Shinjo2009,Zyubin2011,Kravets2015} Further technological progress requires a better understanding of the processes behind magnetic relaxation. A special attention should be paid to spin relaxation found in specific functional nanostructures, not observed in bulk materials.

In bulk ferromagnets, the motion of the magnetization and its damping are well described by the Landau-Lifschitz-Gilbert equation.\cite{Landau1969,Serpico2003,Gilbert2004} The strength of the damping term is scaled by the Gilbert damping constant, $G$, which is a key parameter for spin electronics, since it determines the magnetization switching time and the critical current density in spin-torque based devices.\cite{Katine2000,Petit2007,Brataas2012,Suhl1998}

The Gilbert damping parameter characterizes the energy transfer from the spin subsystem to the lattice.\cite{Gilbert2004} Several microscopic mechanisms intrinsic to ferromagnetic materials, such as phonon drag\cite{Suhl1998} and spin-orbit coupling,\cite{Kambersky1970} have been proposed to account for magnetic relaxation represented by the Gilbert term. There are, however, mechanisms which cannot be described in Gilbert-like form. Two new mechanisms have been the topic of recent discussion regarding magnetization damping in ultrathin films and multilayers: two-magnon scattering\cite{Sparks1964,Zakeri2007} and the spin-pumping effect.\cite{Mizukami2002,Tserkovnyak2002}

Two-magnon scattering is a process, where the magnon of zero wave vector ($k=0$) scatters into degenerate states of magnons having wave vectors $k\neq0$.\cite{Sparks1964} This process requires that the spin-wave dispersion contains degenerate states, and that there are scattering centers in the sample. The geometrical separation of the scattering centers determines the spatial extent of the final magnon states. If long-wavelength spin waves are involved in the relaxation process, defects of the order of several hundreds of nanometers rather than atomic defects act as scattering centers. The existence of two-magnon scattering has been demonstrated in many systems of ferrites.\cite{Hurben1998,Nazarov2003,Mo2005} While in bulk materials this mechanism is well known, it was only recently found to be of importance also for multilayers\cite{Lindner2003,Lenz2006} and ultrathin ferromagnetic films.\cite{McMichael2003,Butera2005,Woltersdorf2004}

In the presence of an interface between a ferromagnetic and a nonmagnetic layer, the spin-pumping effect can cause an increase in the damping constant.\cite{Mizukami2002,Tserkovnyak2002} Excitation of a ferromagnet by a microwave field normally leads to a coherent precession of its spins, which can act as a spin battery injecting through the interlayer interface a pure spin current into the neighboring nonmagnetic layer. Due to the presence of impurity scattering in the system, this spin current can return to the interface, bringing the carried angular momentum back to the precessing spins of the ferromagnetic layer.\cite{Yanson2005} Depending on the parameters of the neighboring layers and interlayer interfaces, a portion of the angular momentum flow will be absorbed by the ferromagnetic layer via various spin-flip relaxation processes. Therefore, the backflow through the nonmagnetic/ferromagnetic interface is always weaker than the direct flow. This imbalance enhances damping of the magnetization precession.\cite{Brataas2012,Tserkovnyak2002,Timopheev2014}

The above spin relaxation effects, being of wide fundamental and applied interest, can be studied most suitably using ferromagnetic resonance (FMR). This powerful method for characterizing magnetic materials relates the measured FMR linewidth to the spin relaxation mechanisms outlined above. For Gilbert-type relaxation, the FMR linewidth (half-width at half maximum) is $\Delta^{\textrm{hwhm}}=\alpha \omega/\gamma$, where $\alpha$ is a dimensionless damping parameter related to $G$ as $\alpha=G/(\gamma M)$, $\omega$ is the angular frequency of exciting field, $\gamma$ is the absolute value of the electron spectroscopic splitting factor, and $M$ is the ferromagnet's magnetization. Damping mechanisms extrinsic to the ferromagnet, such as spin-pumping, result in additional contributions to the measured FMR linewidth and can be deduced by studying effects on the FMR from varying the physical parameters of the multilayer.

We have recently developed magnetic multilayers with temperature-controlled interlayer-exchange coupling -- the so-called \textit{Curie-switch} or \textit{Curie-valve} structures.\cite{Andersson2010a,Andersson2010,Kadigrobov2010} A Curie-switch is a F$_1$/f/F$_2$/AF multilayer where weakly ferromagnetic spacer f is sandwiched between soft ferromagnetic layer F$_1$ and hard ferromagnetic layer F$_2$ exchange-pinned by antiferromagnetic layer AF. Magnetic coupling between F$_1$ and F$_2$ depends on whether temperature $T$ is higher or lower than the Curie temperature of the spacer.\cite{Kadigrobov2012,Kravets2012} As a result, the switching of the magnetic configuration from parallel to antiparallel may be achieved by driving the spacer thermally through its Curie point.\cite{Kravets2015,Kravets2012}

Magnetic relaxation in a Curie-switch has not been fully explored. This work is a study of the FMR properties of F$_1$/f/F$_2$/AF multilayers, aimed at understanding the mechanisms involved and, specifically, the role the interlayer exchange coupling plays in spin relaxation in the system.

\section{Experimental details}

The experiments were carried out on multilayers Py(10)/ Ni$_{54}$Cu$_{46}$($d$)/ Co$_{90}$Fe$_{10}$(5)/Mn$_{80}$Ir$_{20}$(12) [hereinafter -- F$_1$/ Ni$_{54}$Cu$_{46}$($d$)/ F$_2$] with spacer thicknesses $d=$ 3, 4.5, 6, and 9 nm. The numbers in brackets represent layer thicknesses in nanometers. The multilayers were deposited at room temperature on thermally oxidized silicon substrates using magnetron sputtering in an AJA Orion 8-target system. The exchange pinning between the ferromagnetic Co$_{90}$Fe$_{10}$ and antiferromagnetic Mn$_{80}$Ir$_{20}$ layers was set in during deposition using an in-plane magnetic field $H_{\textrm{dep}}\approx0.6$~kOe. Additional fabrication details can be found in Refs.~\onlinecite{Kravets2012,Kravets2014}.

Magnetic properties of the multilayers with a specific spacer composition (Ni$_{54}$Cu$_{46}$) were reported previously.\cite{Kravets2015,Kravets2012,Kravets2014} It was shown that at room temperature, the coupling between F$_1$ and F$_2$ layers strongly depends on the spacer thickness $d$. The increase in $d$ from 3 to 9 nm makes the system transition from a strongly-coupled to a fully exchange-decoupled regime.

FMR measurements were carried out using an X-band ELEXSYS E500 spectrometer equipped with an automatic goniometer. The operating frequency was $\nu=9.44$~GHz. The out-of-plane and in-plane angular dependences of the FMR spectra were studied at room temperature (295 K). The resonance signals from both F$_1$ and F$_2$ were clearly separated in field.

The FMR measurements recorded the first derivative of the microwave absorption by the sample. Each spectrum was fitted by a field derivative of a  Lorentzian function to obtain the relevant resonance field $H_{\textrm{r}i}$ and linewidth $\Delta_{i}=2\Delta_{i}^{\textrm{hwhm}}$ ($i=1,2$ correspond to layers F$_1$, F$_2$).

\section{Theoretical description}
\subsection{Effect of interlayer coupling} \label{A}

Consider a F$_1$/f/F$_2$/AF multilayer, where weakly ferromagnetic spacer f is sandwiched between soft magnetic F$_1$ and hard magnetic F$_2$ exchange-pinned by AF.\cite{Kravets2015,Kravets2014} The thickness's of F$_1$, F$_2$ and f are, respectively, $L_1$, $L_2$ and $d$.

The calculation of the FMR modes will assume that the action of the applied uniform external field does not affect the uniform distribution of the magnetization $\textbf{M}_1$ and $\textbf{M}_2$ in F$_1$ and F$_2$, respectively. In our case of thin layers and strong intralayer exchange interactions, this assumption is well justified.\cite{Kravets2015,Aharoni1996} Spacer f with magnetization $\textbf{m}$ provides a relatively weak coupling between the outer ferromagnetic layers. The aim of this subsection is to determine the effect of this coupling on the FMR linewidth $\Delta_1$.

For a uniform ferromagnetic layer, the energy density consists of magneto-dipole and Zeeman terms. The exchange bias between F$_2$ and AF can be modelled using an effective biasing field, $\textbf{H}_{\textrm{b}}$, acting on the magnetization, $\textbf{M}_2$.\cite{Kravets2015,Nogues2005} Using these notations, the expression for the energy density $w_{i}$ of the $i$-th layer can be written as a sum of the demagnetization term and the terms describing the interaction of the layers' magnetizations with the effective biasing, $\textbf{H}_{\text{b}}$, external quasistatic, $\textbf{H}$, and alternating, $\textbf{h}$, magnetic fields:
\begin{multline}\label{eq:1}
w_i = 2 \pi M_i^2 \cos^2 \theta_i - M_i H_{\textrm{b}i} \cos \varphi_i \sin \theta_i\\
- M_i H \cos (\varphi_i-\varphi_{H})\sin \theta_i - M_{i} h \cos \theta_{i} ,
\end{multline}
where $i=1,2$; $M_{i}$ is the saturation magnetization of the $i$-th layer; $H_{\textrm{b}1}=0$, $H_{\textrm{b}2}=H_{\textrm{b}}$; $H$ is the external quasistatic magnetic field applied in the film plane $xOy$; $h$ is the weak alternating magnetic field applied perpendicular to the film plane; $\varphi_{H}$ is the angle between $\textbf{H}$ and the $Ox$ axis directed along $\textbf{H}_{\textrm{b}}$; $\theta_{i}$ and $\varphi_{i}$ are, respectively, the polar and azimuthal coordinates of the magnetization vector in the $i$-th layer.

In the case of a thin film, its high out-of-plane demagnetization fields prevent the magnetization vector from strongly deviating from the $xOy$ plane. In this case, $\theta_{i}$ can be represented as $\theta_{i}=\pi/2 + \varepsilon_{i}$, where $|\varepsilon_{i}| \ll 1$. This makes it possible to simplify further calculations by expanding the energy density in powers of $\varepsilon_{i}$ and keeping only terms not higher than quadratic in $\varepsilon_{i}$.

The equations of the magnetization dynamics, which take into account the weak coupling between F$_1$ and F$_2$, can be obtained following the procedure described in Ref.~\onlinecite{Kravets2015}. Let us introduce Lagrange function $L$, averaged over two ferromagnetic layers, and dissipative function $\Re$ in the Gilbert form:
\begin{equation}\label{eq:2}
L
 =T-W-\dfrac{4 \pi \Lambda^2 m^2}{2d \left(L_1+L_2 \right)}
  \biggl[
    ( \varphi_1-\varphi_2)^2 + (\varepsilon_1-\varepsilon_2)^2
  \biggr],
\end{equation}
\begin{equation}\label{eq:3}
T
  =\sum^{2}_{i=1}- \dfrac{l_i M_i}{\gamma}\cos \theta_i \dot{\varphi_i}
  \approx \sum^{2}_{i=1} \dfrac{l_i M_i}{\gamma}\varepsilon_i \dot{\varphi_i} \;,
\end{equation}
\begin{multline}\label{eq:4}
W
 =\sum^{2}_{i=1}l_i w_i
 \approx \sum^{2}_{i=1}l_i
  \biggl[
    2\pi M_i^2 \varepsilon_i^2
    - H_{\textrm{b}i} M_i \cos \varphi_i \left( 1-\dfrac{\varepsilon_i^2}{2}\right)
    \\
  -HM_i \left( 1-\dfrac{\varepsilon_i^2}{2}\right)\cos(\varphi_i-\varphi_H)
  \biggr ]\;,
\end{multline}%
\begin{multline}\label{eq:5}
\Re
  =\sum^{2}_{i=1}\dfrac{\alpha_iM_il_i}{2\gamma}\left(\dot{\theta_i^2}+\sin^2   \theta_i \dot{\varphi_i^2}\right)
  \\
 \approx \sum^{2}_{i=1}\dfrac{\alpha_iM_i l_i}{2\gamma}\left(\dot{\varepsilon_i^2}+\dot{\varphi_i^2}\right).
\end{multline}
Here the dot over the angle variables $\theta_i$, $\varphi_i$, and $\varepsilon_i$ means differentiation in time. $T$ and $W$ are the kinetic and potential energy of the system, respectively, $l_i = L_i /(L_1 + L_2)$ is the relative thickness of the $i$-th ferromagnetic layer, $\alpha_i$ is the dissipative constant in the Gilbert form, $\Lambda$ is the magnetic exchange length of the material of the spacer, which is related to the spacer exchange constant $\zeta$ as $\Lambda=\sqrt{\zeta/4\pi}$.\cite{Kravets2015,Abo2013}

The last term in Eq.~(\ref{eq:2}) describes the coupling energy between F$_1$ and F$_2$. Its derivation and the limits of validity are detailed in Ref.~\onlinecite{Kravets2015}.

The equations for the magnetization dynamics in the Lagrange form are:\cite{Landau1969}
\begin{equation}\label{eq:6}
\frac{d}{dt}\dfrac{\partial L}{\partial \dot{\varepsilon_i}}
   =\dfrac{\partial L}{\partial \varepsilon_i} - \dfrac{\partial \Re}{\partial \dot{\varepsilon_i}}\;,
   \qquad
\frac{d}{dt}\dfrac{\partial L}{\partial \dot{\varphi_i}}
   =\dfrac{\partial L}{\partial \varphi_i} - \dfrac{\partial \Re}{\partial \dot{\varphi_i}}\;.
\end{equation}

After substitutions of Eqs.~(\ref{eq:1})--(\ref{eq:5}) into Eq.~(\ref{eq:6}), we obtain:
\begin{multline}\label{eq:7}
\frac{1}{\gamma}\dfrac{d\varepsilon_i}{dt}
    + \frac{\alpha_i}{\gamma}\dfrac{d\varphi_i}{dt}
    + H\sin(\varphi_i-\varphi_{H})
       \\
    + H_{\textrm{b}i} \sin \varphi_i
    - k_{i}(-1)^i (\varphi_1-\varphi_2)=0\;,
\end{multline}
\begin{multline}\label{eq:8}
-\frac{1}{\gamma}\dfrac{d \varphi_i}{dt}
   + \frac{\alpha_i}{\gamma}\dfrac{d \varepsilon_i}{dt}
   + \biggl [
      4 \pi M_i+H\cos(\varphi_i-\varphi_{H})
      \\
   + H_{\textrm{b}i} \cos \varphi_i
    \biggr]
    \varepsilon_{i}
   - k_{i}(-1)^i (\varepsilon_1-\varepsilon_2)
= -h\;,
\end{multline}
where $k_i = 4 \pi \Lambda^2 m^2 /d L_i M_i$ is the effective coupling constant with the dimension of magnetic field, characterizing the exchange from the neighbouring layers on the $i$-th layer.\cite{Kravets2015}

When the alternating magnetic field equals zero ($h = 0$), $\varepsilon_i = 0$, and the equilibrium angles $\varphi_{0i}$ can be determined from the following equations:
\begin{equation}\label{eq:9}
H\sin \left(\varphi_{01}-\varphi_{H}\right)
  + k_1 \left(\varphi_{01}-\varphi_{02}\right)
  = 0\;,
\end{equation}
\begin{equation}\label{eq:10}
H\sin \left(\varphi_{02}
  - \varphi_{H}\right)
  + H_{\textrm{b}} \sin \varphi_{02}
  - k_2 \left(\varphi_{01}-\varphi_{02}\right)
  = 0\;.
\end{equation}

From our earlier work,\cite{Kravets2015,Kravets2014} $H_{\textrm{b}} \sim 300$~Oe, $k_i \sim 150$~Oe, and $H \approx H_{\textrm{r}1} \approx 1200$~Oe, where $H_{\textrm{r}1}$ is the resonance field for F$_1$. As a result, keeping only terms not higher than quadratic in $H_{\textrm{b}}/H$ and $k_i/H$, we can write:
\begin{equation}\label{eq:11}
\varphi_{02} \approx \varphi_{H}-\dfrac{H_{\textrm{b}}}{H}\sin \varphi_{H}\;,
   \qquad
   \varphi_{01} \approx \varphi_{H}\;.
\end{equation}

After writing the angle variables in the form $\varphi_i = \varphi_{0i} + u_i$, where $| u_i | \ll 1$, the linearized system of Eqs.~(\ref{eq:7})--(\ref{eq:8}) can be rewritten as
\begin{widetext}
\begin{equation}\label{eq:12}
\begin{pmatrix}
  iH_{\omega} & i\alpha_1H_{\omega}+H_1 & 0 & -k_1\\
  i\alpha_1H_{\omega}+4 \pi M_1 +H_1 & -iH_{\omega} & -k_1 & 0 \\
  0 & -k_2 & iH_{\omega} & i\alpha_2 H_{\omega}+H_2 \\
  -k_2 & 0 & i\alpha_2 H_{\omega}+4\pi M_2 + H_2 & -iH_{\omega}
\end{pmatrix}
\times
\begin{pmatrix}
\varepsilon_1 \\ u_1 \\ \varepsilon_2 \\ u_2
\end{pmatrix}
=
\begin{pmatrix}
0 \\ -h \\ 0 \\ -h
\end{pmatrix},
\end{equation}
\end{widetext}
where $H_{\omega}=\omega/\gamma$, $H_1=H+k_1$, $H_2=H+H_{\textrm{b}}\cos\varphi_{H}+k_2$.

The coefficients in Eq.~(\ref{eq:12}) were obtained using expansion $\sqrt{H^2+2HH_{\textrm{b}}\cos\varphi_H+H_{\textrm{b}}^2}=H\sqrt{\left(1+H_{\textrm{b}} \cos \varphi_{H}/H \right)^2+\left(H_{\textrm{b}}/H\right)^2 \sin^2 \varphi_H}\approx H+H_{\textrm{b}}\cos\varphi_H$ under the assumption that ($H_{\textrm{b}}/H)^2\ll 1$. For this reason, here and below, the quantitative validity of the calculations is restricted to the terms linear in $H_{\textrm{b}}/H$.

Let us recall that the main task of this subsection consists in determining the angular dependence of the width of the microwave absorption spectrum in the vicinity of the resonance for the free layer, F$_1$. It is evident that this dependence results from the influence of the pinned layer F$_2$ on the free layer through the weakly ferromagnetic spacer. To accomplish the task, it is sufficient to analyze the behavior of the determinant of the matrix in Eq.~(\ref{eq:12}) in the vicinity of $H_{\textrm{r}1}$, and precise analytical determination of $\varepsilon_i(t)$ and $u_i(t)$ is not required.

The absorption intensity $I(H)$ is determined from averaging of the dissipation function over time:\cite{Landau1969}
\begin{equation}\label{eq:13}
I(H)=2\overline{\Re}\sim\dfrac{1}{2}\sum^2_{i=1}\overline{\alpha_i(\dot{\varepsilon_i}\dot{\varepsilon_i}^*+\dot{u_i}\dot{u_i}^*)}\;.
\end{equation}
where the asterisk over the angle variables $\varepsilon_i$ and $u_i$ means conjugate.

The values of $\varepsilon_i$ and $u_i$ are proportional to $1/D$, where  $D=D^{'}+iD^{''}$ is the determinant of the matrix of Eq.~(\ref{eq:12}), and $D^{'}$ and $D^{''}$ are its real and imaginary parts, correspondingly.

It is easy to show that $1/D$ can be represented in the form:
\begin{equation}\label{eq:14}
\dfrac{1}{D}=\dfrac{A}{\delta^{'}+i \delta^{''}}\;,
\end{equation}
where
\begin{equation}\label{eq:15}
A=\dfrac{H_{\omega}^2-H_2(H_2 + 4\pi M_2)+i\alpha_2H_{\omega}(4\pi M_2+2H_2)}{\left[ H_{\omega}^2-H_2(H_2+4\pi M_2)\right]^2},
\end{equation}
\begin{equation}\label{eq:16}
\delta^{'}
   = H_{\omega}^2
   - H(4\pi M_1+H)
   - k_1(4\pi M_1+2H)
   + k_1 k_2 K_0 \;,
\end{equation}
\begin{equation}\label{eq:17}
\delta^{''}
   = H_{\omega}
     \left \{
        \alpha_1 \left[4\pi M_1+2(H+k_1)\right ]
      - \alpha_1 k_1 k_2 K_1
      + \alpha_2 k_1 k_2 K_2
     \right \} \;,
\end{equation}
with
\begin{multline}\label{eq:18}
K_0=\dfrac{4\pi M_1}{H} \left(1+\dfrac{H}{\pi M_2}\right)\\
\times \left(1-\dfrac{M_1}{M_2}+\dfrac{H_{\textrm{b}}}{H_{\textbf{r}1}}\cos\varphi \right)^{-1},
\end{multline}
\begin{equation}\label{eq:19}
K_1= \dfrac{1}{H} \left( 1-\dfrac{M_1}{M_2}+\dfrac{H_{\textrm{b}}}{H_{\textbf{r}1}}\cos\varphi \right)^{-1}\;,
\end{equation}
\begin{multline}\label{eq:20}
K_2 = \dfrac{4 \pi M_1}{H^2} \left[
1+\dfrac{H}{\pi M_2} + \dfrac{H}{4\pi M_1}
\left(1+\dfrac{M_1^2}{M_2^2}\right)
\right]\\
\times \left(1-\dfrac{M_1}{M_2}+\dfrac{H_{\textrm{b}}}{H_{\textbf{r}1}} \cos\varphi \right)^{-2}.
\end{multline}
Here, the terms quadratic in $\alpha_i$ are neglected.

It follows from Eq.~(\ref{eq:13}) that $I(H)\sim A A^*/(\delta^{'2}+\delta^{''2})$. Within a narrow field range in the vicinity of the resonance of F$_1$, sharp changes in the dissipative processes cause changes in $\delta^{'}$ and $\delta^{''}$, while the value of $A$ remains practically unaffected [see Eq.~(\ref{eq:15})] and can be considered as constant.

Resonance conditions for F$_1$ are obtained when  $\delta^{'}=0$. In this case, the absorption intensity approaches its maximal value:
\begin{equation}\label{eq:21}
I(H_{\textrm{r}1})=I_{\textrm{max}}=\textrm{const}/\delta^{''2}.
\end{equation}

When magnetic field $H$ deviates from $H_{\textrm{r}1}$, the absorption intensity decreases, and $I$ becomes one half of $I_{\textrm{max}}$ for $H$ satisfying the following condition:
\begin{equation}\label{eq:22}
\vert \delta^{'} \vert_{H=H_{\textrm{r}1}\pm \Delta_{1}^{\textrm{hwhm}}}=\vert\delta^{''} \vert_{H=H_{\textrm{r}1}}.
\end{equation}

After relatively straightforward but cumbersome transformations based on the condition of Eq.~(\ref{eq:22}), one can obtain the angular dependence of the half-width-at-half-maximum for the absorption intensity curve:
\begin{multline}\label{eq:23}
\dfrac{\Delta_1^{\textrm{hwhm}}}{H_{\omega}} = \alpha_1 \\
-\alpha_1 \dfrac{k_1k_2}{H_{\text{r}1}^2}
\dfrac{\left(1-\dfrac{H_{\textrm{r}1}}{4\pi M_1} \right) \left(1-\dfrac{M_1}{M_2}\right)}
{\left( 1-\dfrac{M_1}{M_2}+\dfrac{H_{\textrm{b}}}{H_{\textrm{r}1}}\cos \varphi\right)^2}\; \\
+\alpha_2 \dfrac{k_1 k_2}{H_{\textrm{r}1}^2}\dfrac{1+\dfrac{H_{\textrm{r}1}}{\pi M_2} - \dfrac{H_{\textrm{r}1}}{4 \pi M_1}
\left(1-\dfrac{M_1^2}{M_2^2} \right)}
{\left(1-\dfrac{M_1}{M_2} + \dfrac{H_{\textrm{b}}}{H_{\textrm{r}1}}\cos \varphi\right)^2} \;.
\end{multline}

In fabricating our samples, condition $L_1 M_1 \approx L_2 M_2$ was kept,\cite{Kravets2012,Kravets2014} which allows reducing the number of independent parameters in the problem by setting $k_1 \approx k_2 \approx k$.

To separate the main factors governing the value of $\Delta_1^{\textrm{hwhm}}$, only terms not higher than quadratic in small parameter $k/H_{\textrm{r1}}$ were kept in Eq.~(\ref{eq:23}). At the same time, the terms which are proportional to
\begin{multline}\label{eq:24}
\left(\dfrac{H_{\textrm{r}1}}{4\pi M_{i}}\right)^2 \leq 0.1\;,\qquad
\left( \dfrac{H_{\textrm{b}}}{H_{\textrm{r}1}}\right)^2 \leq 0.1 \;, \\
\dfrac{H_{\textrm{b}}}{4 \pi M_i} = \dfrac{H_{\textrm{b}}}{H_{\textrm{r}1}} \dfrac{H_{\textrm{r}1}}{4 \pi M_i} \leq 0.1 \;,
\end{multline}
were neglected (the corresponding values were estimated based on the results of Refs.~\onlinecite{Kravets2015,Kravets2012,Kravets2014}).

It is noteworthy that the right hand side of Eq.~(\ref{eq:23}) does not contain terms linear in $k_i$: the angular dependence in $\Delta_1^{\textrm{hwhm}}$ first appears via a product of the coupling constants, $k_1$ and $k_2$. Such kind of $\Delta_1^{\textrm{hwhm}}$ vs $k_i$ dependence reflects complex cross-excitation processes between the outer ferromagnetic layers, F$_1$ and F$_2$. Due to the non-negligible coupling between the layers, the magnetization of F$_2$ is ``dragged" into oscillations by the resonant precession in F$_1$. A simultaneous, but much weaker inverse effect occurs: the magnetization of F$_1$ experiences an exchange-drag from the precession in F$_2$. It is such kind of cross-excitations that affects the relaxation processes in F$_1$ and, depending on the parameters of both ferromagnetic layers coupled via the spacer, this either weakens or enhances the total damping.

Let us consider the situation in the vicinity of the resonance in F$_1$. Compared to an isolated F$_1$, where all microwave energy would be stored within the layer, the flow of the energy in the coupled F$_1$--F$_2$ system divides into two channels: a portion remains stored in F$_1$ while the remaining precessional energy leaves outwards and later returns via the above cross-excitation processes. One should keep in mind that there is an additional energy gain in the second channel, which originates from the excitation of F$_2$ by the external magnetic field. The total energy losses in the system are governed by the processes in both channels. The energy dissipation in the first channel is determined by the intrinsic relaxation mechanisms in F$_1$, but the energy flow through the second channel depends on the relationship between the processes of energy loss and gain in F$_2$. If there is no damping in F$_2$ ($\alpha_2=0$) or it is relatively weak ($\alpha_2 < \alpha_1$), the energy losses in the second channel will respectively be zero or small (in comparison with the losses in F$_1$). Accounting for the additional energy gain due to the excitation of F$_2$ by the external magnetic field, the total energy losses in the coupled F$_1$--F$_2$ system will be smaller than the losses in an isolated F$_1$ and, therefore, the total effective damping parameter of the F$_1$ layer will be smaller than $\alpha_1$. On the contrary, if the energy dissipation in F$_2$ is relatively strong ($\alpha_2 > \alpha_1$), the energy losses in the second channel will be enhanced, and the total effective damping parameter will be greater than $\alpha_1$.

The in-plane anisotropy of $\Delta_1^{\textrm{hwhm}}$ originates from the angle dependence of the denominator in the second and third terms of the right hand side part of Eq.~(\ref{eq:23}). A close look at the denominator reveals that it represents an approximate form of the difference between the resonance fields of F$_1$ and F$_2$. This reflects the fact that the efficiency of the cross-excitation processes in the coupled F$_1$--F$_2$ system depends not only on the coupling constants $k_1$ and $k_2$, but also on the difference between $H_{\textrm{r1}}$ and $H_{\textrm{r2}}$: the smaller the difference, the more efficient the processes. In nanostructures of the spin-valve type, the effect of the exchange bias field, $\textbf{H}_{\textrm{b}}$, is strong in-plane unidirectional anisotropy of $H_{\textrm{r2}}$, with $H_{\textrm{r2}}$ maximally approaching $H_{\textrm{r1}}$ when the external magnetic field is directed opposite to $\textbf{H}_{\textrm{b}}$ ($\varphi=180^{\circ}$).\cite{Kravets2015} As a result, the cross-excitation processes become most efficient at $\varphi=180^{\circ}$ and the above (anti)damping contributions to $\Delta_1^{\textrm{hwhm}}$ from F$_2$ become most pronounced at this angle.

\begin{figure}
\centering
\includegraphics[width=\linewidth]{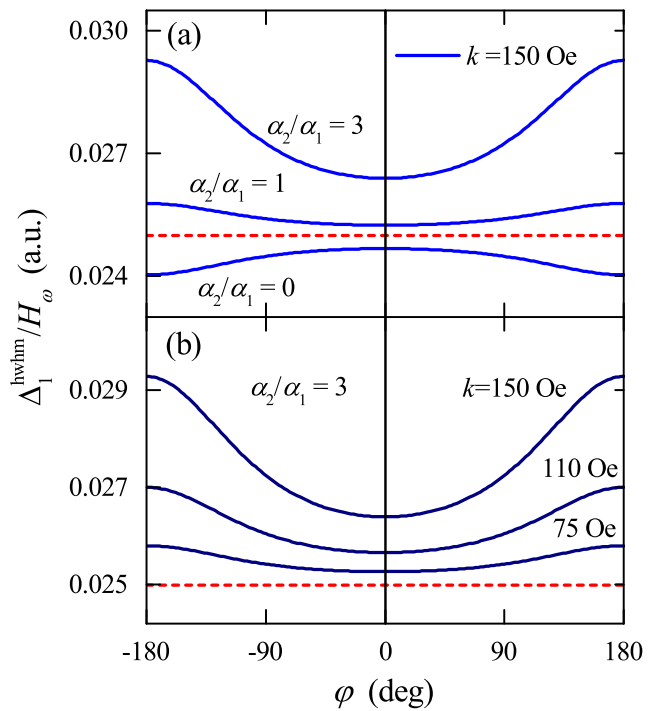}
\caption{In-plane angle dependence of normalized FMR linewidth $\Delta_1^{\textrm{hwhm}}/H_{\omega}$ for different ratios of $\alpha_2 / \alpha_1$ (a) and different $k$ (b). Dotted lines are for a would-be isolated F$_1$ layer, whose damping parameter is $\alpha_1$.}
\label{figure1}
\end{figure}

The above effects are illustrated in Fig.~\ref{figure1}. Model calculations are carried out with the use of Eq.~(\ref{eq:23}). Figure~\ref{figure1}(a) shows the in-plane angle dependencies of the normalized FMR linewidth $\Delta_1^{\textrm{hwhm}}/H_w$ for different ratios of $\alpha_2 / \alpha_1$ for the case of a moderate interlayer coupling ($k=150$ Oe). Dotted line represents the same dependence for an isolated F$_1$ layer whose damping parameter is  $\alpha_1$. It is seen that the value of $\alpha_2$ strongly affects the character of $\Delta_1^{\textrm{hwhm}}/H_w$ vs $\varphi$ dependencies. For the case of weak energy dissipation in F$_2$ ($\alpha_2 =0$), the total energy losses in the coupled F$_1$--F$_2$ system are smaller than the intrinsic losses in F$_1$ so the total effective damping parameter of F$_1$ is smaller than $\alpha_1$. On the contrary, if the energy dissipation in F$_2$ is relatively strong ($\alpha_2 > \alpha_1$), the total effective damping parameter is greater than $\alpha_1$. The increase in $\alpha_2$ results in both an overall increase in $\Delta_1^{\textrm{hwhm}}$ and an enhancement of its in-plane anisotropy. In all cases, the contribution to $\Delta_1^{\textrm{hwhm}}$, induced by the interlayer coupling, is minimal at $\varphi=0$ and maximal at $180^{\circ}$. As described above, the difference between $H_{\text{r1}}$ and $H_{\textrm{r2}}$, and hence the cross-excitation processes, achieve opposite extrema at these specific values of the in-plane angle, which is in good agreement with the experimentally observed behavior (see below).

Figure~\ref{figure1}(b) illustrates the evolution of $\Delta_1^{\textrm{hwhm}}/H_w$ vs $\varphi$ with changes in the coupling constant $k$. The increase in $k$ leads to both an overall increase in $\Delta_1^{\textrm{hwhm}}$ and an enhancement of the in-plane anisotropy of $\Delta_1^{\textrm{hwhm}}$, as a result of stronger cross-excitation processes in this stronger exchange-coupling case.

\subsection{Effect of dispersion in local fields} \label{B}
	
To correctly analyze various contributions to the FMR linewidth, one should take into account the broadening of the linewidth due to fluctuations of the magnetic parameters in the structure, always present on the experiment.

In general, FMR in a finite-size ferromagnet is governed by the effective magnetization, $\textbf{M}_{\textrm{eff}}$, which includes contributions from the spontaneous magnetization, $\textbf{M}$, influenced by the local shape, strain, and crystalline anisotropy.\cite{Gurevich1996} As shown in Ref.~\onlinecite{Kravets2015}, the resonance in F$_1$ and F$_2$ is governed by the in-plane contributions from the uniaxial, unidirectional, and shape anisotropy, relevant for each of the ferromagnetic layers. The weak uniaxial anisotropy is likely due to the applied field during the multilayer deposition. The unidirectional anisotropy is due to the biasing field, $\textbf{H}_{\textrm{b}}$, acting on $\textbf{M}_2$ and, via the interlayer coupling, on $\textbf{M}_1$.\cite{Kravets2015}

Let us consider the effect of such dispersion in the effective magnetization on the FMR linewidth. Restricting our consideration to the above anisotropy contributions, we can write:
\begin{equation}\label{eq:25}
H_{\textrm{r}}=f(\textbf{M},\textbf{H}_{\textrm{ua}}, \textbf{H}_{\textrm{ud}}),
\end{equation}
where $H_{\textrm{r}}$ is the magnitude of the resonance field, $\textbf{H}_{\textrm{ua}}$ and $\textbf{H}_{\textrm{ud}}$ are the uniaxial and unidirectional anisotropy fields, respectively. In this case, the inhomogeneous linewidth broadening due to fluctuations in the magnitudes and directions of $\textbf{M}$, $\textbf{H}_{\textrm{ua}}$ and $\textbf{H}_{\textrm{ud}}$ can be written as
\begin{equation}\label{eq:26}
\Delta^{\textrm{inhom}}=\Delta_{\textbf{M}}+\Delta_{\textbf{H}_{\textrm{ua}}}+\Delta_{\textbf{H}_{\textrm{ud}}} \;,
\end{equation}
where
\begin{multline}\label{eq:27}
\Delta_{\textbf{M}}
=\Delta_{M} + \Delta_{\theta} + \Delta_{\varphi} \\
=\left|\dfrac{\partial H_{\textrm{r}}}{\partial M}\right| \delta M
+\left|\dfrac{\partial H_{\textrm{r}}}{\partial \theta }\right| \delta \theta
+\left|\dfrac{\partial H_{\textrm{r}}}{\partial \varphi} \right| \delta \varphi \;,
\end{multline}
\begin{multline}\label{eq:28}
\Delta_{\textbf{H}_{\textrm{ua}}}
=\Delta_{H\textrm{ua}}+\Delta_{\theta\textrm{ua}}+\Delta_{\varphi\textrm{ua}} \\
=\left|\dfrac{\partial H_{\textrm{r}}}{\partial H_{\textrm{ua}}} \right| \delta H_{\textrm{ua}}
+\left|\dfrac{\partial H_{\textrm{r}}}{\partial \theta_{\textrm{ua}}}\right| \delta \theta_{\textrm{ua}}
+\left|\dfrac{\partial H_{\textrm{r}}}{\partial \varphi_{\textrm{ua}}} \right| \delta \varphi_{\textrm{ua}}\;,
\end{multline}
\begin{multline}\label{eq:29}
\Delta_{\textbf{H}_{\text{ud}}}
=\Delta_{H\textrm{ud}}+\Delta_{\theta\textrm{ud}}+\Delta_{\varphi\textrm{ud}} \\
=\left|\dfrac{\partial H_{\textrm{r}}}{\partial H_{\textrm{ud}}} \right| \delta H_{\textrm{ud}}
+\left|\dfrac{\partial H_{\textrm{r}}}{\partial \theta_{\textrm{ud}}}\right| \delta \theta_{\textrm{ud}}
+\left|\dfrac{\partial H_{\textrm{r}}}{\partial \varphi_{\textrm{ud}}} \right| \delta \varphi_{\textrm{ud}}\;.
\end{multline}

Here, $\Delta_{\textbf{M}}$, $\Delta_{\textbf{H}_{\textrm{ua}}}$, and $\Delta_{\textbf{H}_{\textrm{ud}}}$ are the contributions to the FMR linewidth caused by the dispersion in $\textbf{M}$, $\textbf{H}_{\textrm{ua}}$ and $\textbf{H}_{\textrm{ud}}$, respectively, which in Eqs.~(\ref{eq:27})--(\ref{eq:29}) are expressed through the corresponding distributions in magnitudes ($\delta M$, $\delta H_{\textrm{ua}}$, and $\delta H_{\textrm{ud}}$) as well as polar ($\delta\theta$, $\delta\theta_{\textrm{ua}}$, $\delta\theta_{\textrm{ud}}$) and azimuthal ($\delta\varphi$, $\delta\varphi_{\textrm{ua}}$, $\delta \varphi_{\textrm{ud}}$) angles characterizing these vectors.

Based on the analysis of the partial derivatives of $H_{\textrm{r}}$, which are contained in Eqs.~(\ref{eq:27})--(\ref{eq:29}), it is possible to separate each contribution to $\Delta^{\textrm{inhom}}$ by analyzing the out-of-plane and in-plane behavior of the FMR linewidth.\cite{Zakeri2007, Woltersdorf2004,Mizukami2001} For example, when out-of-plane measurements are carried out, the azimuthal angle is constant, which means that all contributions containing derivatives in azimuthal angles are constant. For in-plane measurements, on the other hand, all contributions containing derivatives in polar angles are constant. One should also take into account that there are points, where some of the partial derivatives found in Eqs.~(\ref{eq:27})--(\ref{eq:29}) vanish, making it possible to separate the remaining contributions.

\begin{figure}
\centering
\includegraphics[width=\linewidth]{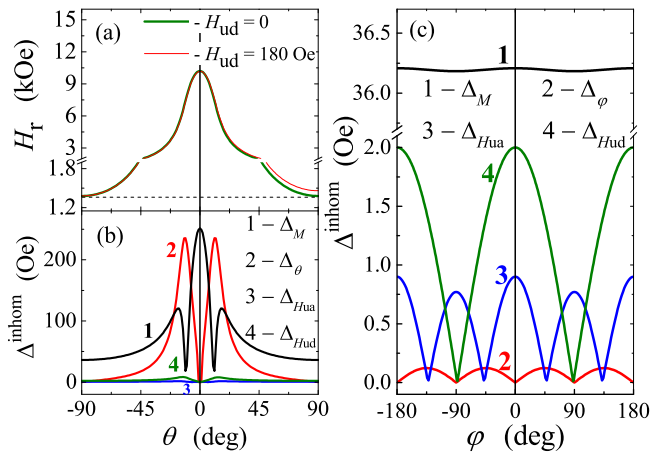}
\caption{(a) Simulated out-of-plane angle dependence of the resonance field, $H_{\textrm{r}}$, for a thin ferromagnetic layer with $M=520$~emu/cm$^3$, $H_{\textrm{ua}}=5$~Oe and $H_{\textrm{ud}}=0$ or 180 Oe. (b) Calculated contributions to out-of-plane $\Delta^{\textrm{inhom}}$ due to fluctuations in magnitude (line 1) and direction (line 2) of spontaneous magnetization, magnitude of in-plane uniaxial (line 3) and unidirectional (line 4) anisotropy fields. Calculations in (a) and (b) were carried out for the $xOz$ plane, where the $Ox$ axis coincides with $\textbf{H}_{\textrm{b}}$ ($\varphi=0$ for negative $\theta$ and $\pm 180^{\circ}$ for positive $\theta$). (c) Calculated contributions to in-plane $\Delta^{\textrm{inhom}}$ due to fluctuations in magnitude (line 1) and direction (line 2) of spontaneous magnetization, magnitude of in-plane uniaxial (line 3) and unidirectional (line 4) anisotropy fields. The distributions in $\textbf{M}$, $\textbf{H}_{\textrm{ua}}$ and $\textbf{H}_{\textrm{ud}}$ were chosen to be: $\delta M=5\%$, $\delta\theta=\delta\varphi=1^{\circ}$, $\delta H_{\textrm{ua}}=15\%$, $\delta H_{\textrm{ud}}=5\%$. }
\label{figure2}
\end{figure}

To analyze the various contributions to $\Delta^{\textrm{inhom}}$, we have simulated the out-of-plane and in-plane angle behavior of the resonance field for a thin ferromagnetic layer, which simultaneously displays in-plane uniaxial and unidirectional anisotropy (the easy axes coincide), and numerically calculated all partial derivatives in Eqs.~(\ref{eq:27})--(\ref{eq:29}). Parameters $\textbf{M}$, $\textbf{H}_{\textrm{ua}}$ and $\textbf{H}_{\textrm{ud}}$ were chosen to be close in values to those observed for Py as F$_1$ and the spacer thickness $d=3$~nm.\cite{Kravets2015} The results of the calculations are shown in Fig.~\ref{figure2}.

It is clear from Fig.~\ref{figure2} that the behavior in the out-of-plane geometry is very sensitive to the scatter in $\theta$ and $M$, and practically insensitive to the anisotropy parameters. On the other hand, the in-plane behavior provides information on the scatter in both magnitude and orientation of the anisotropy fields (for both uniaxial and unidirectional contributions), and is almost insensitive to the scatter in $\theta$ and $M$. Worth to note is that for the out-of-plane geometry, $\Delta^{\textrm{inhom}}\vert_{\theta=0}=\Delta^{\textrm{inhom}}\vert_{\theta=\pm 90^{\circ}}$ in all cases, except for the case where there is a substantial scatter in $M$.

We point out a peculiar result, important for further analysis of the experimental data, namely that $\Delta_{H\textrm{ud}} \vert_{\theta=-90^{\circ}} =\Delta_{H\textrm{ud}}\vert_{\theta=+90^{\circ}}$, while the resonance field at $\theta=-90^{\circ}$ differs from that at $\theta=+90^{\circ}$.

\subsection{Two-magnon scattering} \label{C}
	
The nature of the dispersion relation of spin waves in ultrathin ferromagnets with in-plane magnetization is such that there can be spin-wave modes of finite wave vector degenerate in frequency with the FMR-exited mode.\cite{Sparks1964,Zakeri2007} In the ideal case of a non-dissipative material, all spin wave modes are independent, decoupled normal modes of the system, so the FMR mode does interact with the finite wave-vector modes of the same frequency. However, if defects of random spatial character are present, they can scatter the zero wave-vector FMR spin-wave into a manifold of degenerate modes.\cite{Sparks1964,Zakeri2007,Lindner2003,Lenz2006,McMichael2003,Butera2005,Woltersdorf2004} This can be viewed as a dephasing contribution to the linewidth, in the language of spin-resonance physics.

The two-magnon mechanism is allowed when the magnetization lies in the film plane or slightly deviates from it, and forbidden when the magnetization is perpendicular to the film plane.\cite{Zakeri2007,Erickson1992,Landeros2008} Thus, inequality $\Delta \vert_{\theta=0} < \Delta \vert_{\theta= \pm 90^{\circ}}$ indicates that two-magnon scattering is potentially relevant for the extrinsic magnetization damping in our case.\cite{Lindner2003}

As a rule, $\Delta \vert_{\theta=0} < \Delta \vert_{\theta= \pm 90^{\circ}}$ implies that two-magnon scattering plays a negligible role in magnetization relaxation. However, there are specific cases when this damping mechanism displays strong in-plane anisotropy.\cite{Lindner2003,Woltersdorf2004,Arias2000,McMichael2002,Twisselmann2003} Since the two-magnon scattering matrix includes elements proportional to the components of the Fourier transform of the spatial distribution of magnetic inhomogeneities, the in-plane anisotropy is expected to be pronounced for the case of oriented extended inhomogeneities, such as rectangular networks of line defects,\cite{Lindner2003,Woltersdorf2004} parallel steps\cite{Arias2000} or grooves,\cite{McMichael2002} etc. One cannot exclude the formation of oriented networks of defects or other inhomogeneous entities in nanostructures deposited under relatively high external magnetic field, such as ours.

\subsection{Spin pumping} \label{D}
	
In the case where the spin diffusion length $L_{\textrm{s}}$ of the spacer is smaller than its thickness, the spin current injected by F$_1$ into the spacer is strongly reduced. For a Curie-switch this means that: (i) the variation of the spacer thickness should not affect the relaxation of $\textbf{M}_1$ through the mechanism of spin pumping and (ii) the presence of the pinned layer F$_2$ should not contribute to the anisotropic damping in F$_1$ through the same mechanism.

In nonmagnetic metals, $L_{\textrm{s}}$ is of the order of tens or hundreds of nanometers and in some cases may reach a few micrometers.\cite{Baas2007} Addition of impurities or rising temperature reduce $L_{\textrm{s}}$. In magnetically ordered materials, especially in alloys, $L_{\textrm{s}}$ is strongly reduced compared to nonmagnetic metals. For example, at 4.2 K, spin diffusion length is about 21 nm for Ni, $\sim$8.5 nm for Fe, and $\sim$5.5 nm for Ni$_{84}$Fe$_{16}$.\cite{Baas2007,Moreau2007} At room temperature, $L_{\textrm{s}}$ for Ni$_{84}$Fe$_{16}$ is almost 2 times shorter, about 3 nm. \cite{Baas2007}

For Cu-Ni alloys, rough estimates of $L_{\textrm{s}}$ can be made based on the data of Ref.~\onlinecite{Baas2007}. With the increase in Ni content from 6.9\% to 22.7\%, the spin diffusion length decreases from $\sim$23 to $\sim$7.5 nm at 4.2 K. It is expected that $L_{\textrm{s}}$ will be further reduced with the increase in Ni concentration. It is also expected that the temperature rise to 295 K will additionally reduce $L_{\textrm{s}}$ by 1.5--2 times, likely making it smaller than 3 nm (the minimal spacer thickness in this study) for Ni$_{54}$Cu$_{46}$ at room temperature. For this reason, the contribution of the spin pumping mechanism to the anisotropic damping in F$_1$ will be neglected for the samples in this study.

\section{Experimental results and discussion}
	
\begin{figure}
\centering
\includegraphics[width=\linewidth]{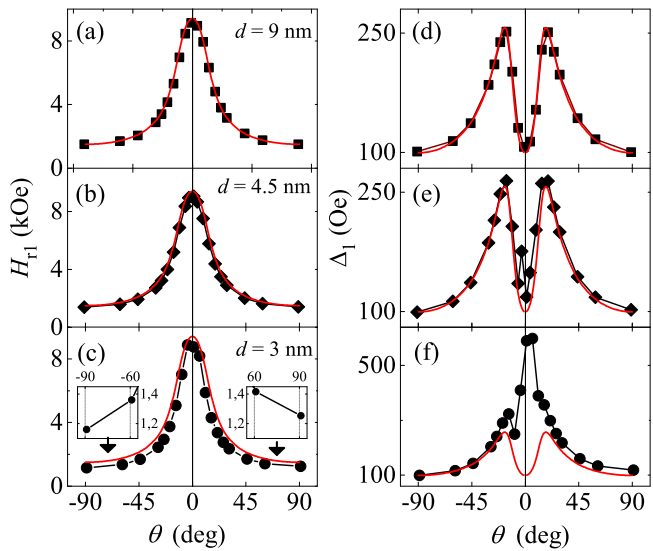}
\caption{Out-of-plane angular dependences of F$_1$ resonance fields (a)--(c) and linewidths (d)--(f) for F$_1$/Ni$_{54}$Cu$_{46}(d)$/F$_2$ multilayers with different spacer thicknesses. Solid red lines in all panels show simulated angular dependence for $d=9$ nm, as described in the text. Insets in panel (c) present enlarged views of $H_{\textrm{r1}}(\theta)$ near $\theta=\pm 90^{\circ}$.}
\label{figure3}
\end{figure}

Figures \ref{figure3}(a)--\ref{figure3}(c) show the measured resonance field for F$_1$ versus the out-of-plane angle for multilayers F$_1$/Ni$_{54}$Cu$_{46}(d)$/F$_2$ with $d=$ 9, 4.5 and 3 nm. The measurements were carried out in the $xOz$ plane, where the $Ox$ axis coincides with $\textbf{H}_{\textrm{b}}$ ($\varphi=0^{\circ}$ for negative $\theta$ and $\pm 180^{\circ}$ for positive $\theta$) and the $Oz$ axis is the normal to the film plane.

For the sample with $d=9$ nm, the behavior of $H_{\textrm{r1}}(\theta)$ is typical of a single permalloy film. The data are quantitatively well described using the Smit-Beljers-Suhl formalism (solid lines in Fig.~\ref{figure3} are the simulated angular behavior with $M_{1}^{\textrm{eff}}=540$ emu/cm$^3$).\cite{Suhl1955,Smit1955} A decrease in $d$ does not lead to noticeable changes in $H_{\textrm{r}1}(\theta)$, but makes the emergence of unidirectional anisotropy evident: $H_{\textrm{r1}}(+90^{\circ})$ becomes greater than $H_{\textrm{r1}}(-90^{\circ})$, and the difference between $H_{\textrm{r1}}(+90^{\circ})$ and $H_{\text{r1}}(-90^{\circ})$ grows as $d$ decreases [see insets to Fig.~\ref{figure3}(c)]. We have previously shown that the unidirectional anisotropy originates from the biasing field $\textbf{H}_{\textrm{b}}$ acting on $\textbf{M}_2$, which in turn transmits through the spacer and affects the FMR behavior of $\textbf{M}_1$.\cite{Kravets2015}

The out-of-plane angular dependences of the FMR linewidth for the same multilayers are shown in Figs.~\ref{figure3}(d)--\ref{figure3}(f). Since the scatter in the magnetic parameters of the decoupled ferromagnetic layer gives different contributions to the linewidth versus angle dependence [see Fig.~\ref{figure2}(b)], a detailed analysis of the measured $\Delta_{1}(\theta)$ curves makes it possible to separate the various local dispersion contributions, as well as the homogeneous contribution.

For the sample with $d=9$ nm, the angular variation in the FMR linewidth $\Delta_1(\theta)$ is well described by a homogeneous term within the Smit-Beljers-Suhl formalism.\cite{Smit1955,Suhl1955} This means that the scatter in both the magnitude and orientation of the magnetization in F$_1$ is negligibly small. The equality $\Delta_1(0^{\circ})=\Delta_1(\pm 90^{\circ})$ serves as an additional confirmation of the fact that $\delta M_{\textrm{Py}}$ is negligible [see line 1 in Fig.~\ref{figure2}(b)]. The same equality also implies that two-magnon scattering plays a negligible role, at least in the $xOz$ plane, which includes only two in-plane directions ($\varphi=0^{\circ}$ and $\pm 180^{\circ}$). This, however, does not exclude that two-magnon scattering can contribute to $\Delta_1$ at other in-plane angles (see Subsection~\ref{C} and discussion below for a more detailed analysis of this mechanism).

For stronger interlayer exchange-coupling, the shape of the $\Delta_{1}(\theta)$ curves shows strong distortions. First, the enhancement of the interlayer interaction leads to an increase in $\Delta_1$ within a relatively narrow range of angles near $\theta=0^{\circ}$. Second, the $\Delta_1(\theta)$ dependence transforms from being symmetric to asymmetric: $\Delta_1(+\vert \theta \vert)$ becomes greater than $\Delta_1(-\vert \theta \vert)$.

It is worth noting that the first effect cannot be caused by the increase of dispersion in the F$_1$ magnetization as that would substantially increase the linewidth not only for $\theta$ near zero, but also for other $\theta$ values [in particular, for $\theta=\pm 90^{\circ}$, see line 1 in Fig.~\ref{figure2}(b)], which is not observed in our experiments. Increased linewidth values within a relatively narrow angle range near $\theta=0^{\circ}$ were observed in Py/Cu and Py/CuAu multilayers at certain values of the spacer thickness ($d_{\textrm{Cu}}=3$ nm or $d_{\textrm{CuAu}}=1.4$ nm),\cite{Dubowik2002} but no particular explanation was suggested for this effect.

As regards the second effect, namely the $\Delta_1(\theta)$ dependence becoming asymmetric, two remarks are in order. (i) This effect is unlikely to result from the scatter in the unidirectional anisotropy fields, since, in spite of the \textit{asymmetric} character of the $H_{\textrm{r1}}(\theta)$, the $H_{\textrm{ud}}$ dispersion contributes \textit{symmetrically} to the linewidth vs $\theta$ dependence [see line 4 in Fig.~\ref{figure2}(b)]. (ii) Asymmetry is expected as a result of the enhanced interlayer coupling [see Eq.~(\ref{eq:23}) and Fig.~\ref{figure1}]. As detailed in Subsection~\ref{A}, a complementary and more detailed information on this effect can be obtained from the in-plane FMR measurements.

\begin{figure}
\centering
\includegraphics[width=\linewidth]{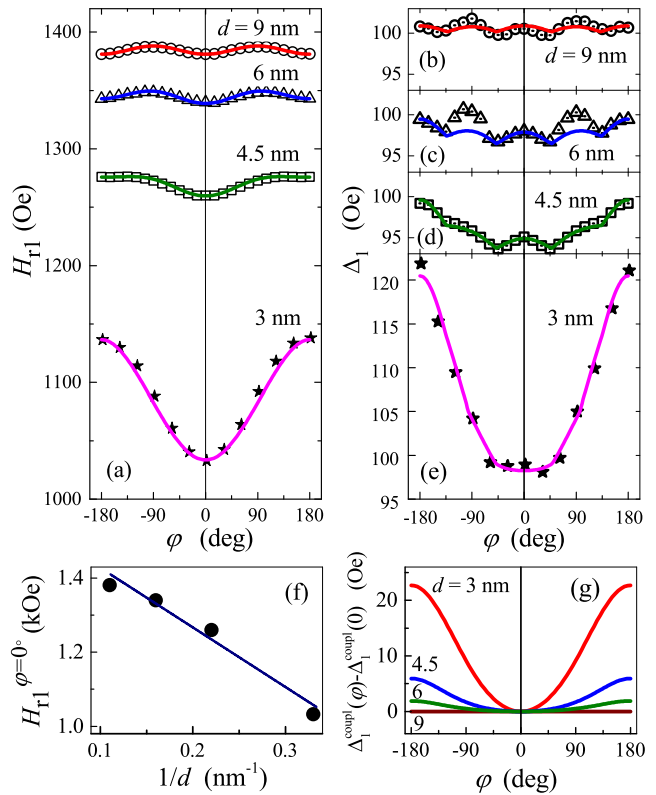}
\caption{In-plane angle dependence of the F$_1$ resonance field (a) and linewidth (b)--(e) for  F$_1$/Ni$_{54}$Cu$_{46}(d)$/F$_2$ multilayers with different spacer thicknesses $d$. Solid lines show the simulated behavior for each sample, obtained as described in the text. (f) Resonance field $H_{\textrm{r1}}$ measured at $\varphi=0^{\circ}$ versus $1/d$. (g) Anisotropic contribution to the FMR linewidth due to the interlayer coupling. Superscript ``coupl" means that only this term from the simulated curves of (b)--(e) is taken into account [described by Eq.~(\ref{eq:23})]. }
\label{figure4}
\end{figure}

Figure~\ref{figure4}(a) shows the in-plane angle dependence of the F$_1$ resonance field for multilayers F$_1$/Ni$_{54}$Cu$_{46}(d)$/F$_2$ with $d=9$, 6, 4.5, and 3 nm. The $H_{\textrm{r1}}(\varphi)$ dependence for the sample with the thickest spacer ($d=9$ nm) provides evidence of a weak in-plane uniaxial anisotropy ($H_{\text{ua}}\approx 5$ Oe). This contribution is a consequence of the application of external magnetic field during the film deposition, as follows from our tests on Py films grown with and without biasing field. The uniaxial contribution to the anisotropy of the Py layer is found in all of the samples with pinned bottom magnetic layers (deposited in field). As $d$ decreases, an additional unidirectional contribution becomes evident and dominates for $d$ thinner than 4.5 nm. This contribution is enhanced for stronger interlayer coupling (see above and also Ref.~\onlinecite{Kravets2015}).

The measured $H_{\textrm{r1}}(\varphi)$ were quantitatively analyzed using the formalism developed in Subsection~\ref{A} and Ref.~\onlinecite{Kravets2015}. Solid lines in Fig.~\ref{figure4}(a) are the calculated angular dependence, for which the parameters were either taken from the previous work ($\Lambda$, $M_1^{\textrm{eff}}$, $M_2^{\textrm{eff}}$),\cite{Kravets2015,Kravets2014} or obtained from fitting the above $H_{\textrm{r1}}(\varphi)$ data with theoretical $m, H_{\textrm{b}}$ and $H_{\textrm{ua}}$. All these key parameters are presented in Table~\ref{table1}. It is important to note that the obtained $m$ and $H_{\textrm{b}}$ are in good agreement with the results reported earlier.\cite{Kravets2015}
\begin{table*}
\caption{\label{table1}Physical parameters obtained from fitting the experimental data on Curie-switch multilayers: $\Lambda$ is the exchange length of the spacer, $M_1^{\textrm{eff}}$ and $M_2^{\textrm{eff}}$ -- effective magnetization of F$_1$ and F$_2$, $m$ -- effective magnetization of the spacer, $H_{\textrm{ua}}$ -- uniaxial anisotropy field of F$_1$ layer, $H_{\textrm{b}}$ -- biasing field acting on $M_2$, $k$ -- interlayer coupling constant, and $\Delta_1^*=\Delta_1(\pm 180^{\circ}) -\Delta_1(0^{\circ})$ -- difference in $\Delta_1$ values measured along and opposite to $\textbf{H}_{\textrm{b}}$.}

\begin{ruledtabular}
\begin{tabular}{ccccccccc}
$d$ (nm) &$\Lambda$ (nm) &$M_{1}^{\textrm{eff}}$ (emu/cm$^3$) &$M_2^{\textrm{eff}}$ (emu/cm$^3$)& $m$ (emu/cm$^3$) &$H_{\textrm{ua}}$ (Oe) &$H_{\textrm{b}}$ (Oe) &$k$ (Oe)& $\Delta_1^*$ (Oe)\\
\hline
3   &11 &520 &1590 &84 &5 &140 &690 & 23  \\
4.5 &11 &520 &1590 &53 &5 &240 &180 & 6 \\
6   &11 &520 &1590 &43 &5 &270 &90  & 2 \\
9   &-  &540 &-    &-  &5 &-   &1   & $\sim$ 0.2 \\
\end{tabular}
\end{ruledtabular}
\end{table*}

Fig.~\ref{figure4}(a) shows that stronger interlayer coupling induces unidirectional anisotropy in F$_1$ as well as overall lowers the resonance field, $H_{\textrm{r1}}$. This behavior can be readily understood within the approach developed in Subsection~\ref{A}. Following Eq.~(\ref{eq:14}), the resonance conditions for F$_1$ are fulfilled when $\delta^{'}$ equals zero. The analysis of the expression for $\delta^{'}$ [see Eq.~(\ref{eq:16})] shows that, to the first order, $H_{\textrm{r1}}$ is a linear function of $k$. Since in our case the coupling constant is inversely proportional to the spacer thickness, $H_{\textrm{r1}}$ should be a linear function of $1/d$. Figure~\ref{figure4}(f) presents the experimentally obtained $H_{\textrm{r1}}$ vs $1/d$ dependence, with the data points falling quite well on a straight line. This fact, along with the good agreement between the experimental and simulated $H_{\textrm{r1}}(\varphi)$, in addition to the experiment-fitting results reported in Ref.~\onlinecite{Kravets2015}, points to the validity of the theory developed herein for the description of the effect of the interlayer exchange-coupling in a Curie-switch.

Further, the parameters obtained from the analysis of $H_{\textrm{r1}}(\varphi)$ allowed us to employ the developed theoretical approach to characterize the angular dependences of the FMR linewidth, $\Delta_1(\varphi)$ [Figs.~\ref{figure4}(b)--\ref{figure4}(e)]. Let us first concentrate on $\Delta_1(\varphi)$ for the sample with $d=9$ nm [Fig.~\ref{figure4}(b)]. For this case of a weak interlayer coupling, the homogeneous contribution to the linewidth does not display any noticeable angular dependence, as shown in Fig.~\ref{figure1}. Thus, the clearly visible variation with a 90-degree periodicity, seen in fact in all samples, is likely to due to local inhomogeneities, namely from scatter in $H_{\textrm{ua}}$ values [line 3 in Fig.~\ref{figure2}(c)]. The solid line in Fig.~\ref{figure4}(b) is the simulated $\Delta_1(\varphi)$ dependence, taking into account only two contributions: $2\Delta_1^{\textrm{hwhm}}$ from Eq.~(\ref{eq:23}) and $\Delta_{H_{\textrm{ua}}}$. The good agreement between the experimental and simulated $\Delta_1(\varphi)$ data shows that this effect is due mainly to a scatter in the values of the uniaxial anisotropy field ($\delta H_{\textrm{ua}} \leq 0.2 H_{\textrm{ua}}$) in the soft ferromagnetic layer.

As $d$ decreases, the shape of $\Delta_1(\varphi)$ undergoes a significant transformation, resulting, in particular, in a much larger difference between $\Delta_1(\pm 180^{\circ})$ and $\Delta_1(0^{\circ})$ (reaching 23 Oe for $d=$ 3 nm). Such changes cannot be ascribed to any dispersion-type contribution or two-magnon scattering. The use of Eq.~(\ref{eq:23}), on the other hand, makes it possible to well describe this behavior in $\Delta_1(\varphi)$ by taking into account the effect of the interlayer coupling.

In addition to the above mentioned contributions to the linewidth, a scatter in the magnitude of the unidirectional anisotropy, $\Delta_{H\textrm{ud}}$ ($\delta H_{\textrm{ud}}<0.05 k $) was taken into account in the calculations [line 4 in Fig.~\ref{figure2}(c)]. This contribution is clearly visible in the data for the sample with $d=3$ nm as a plateau in the range of angles $-45^{\circ}\leq \varphi \leq 45^{\circ}$.

Figures~\ref{figure4}(b)--\ref{figure4}(d) illustrate the good agreement between the measured and modelled $\Delta_1(\varphi)$ for all of the studied samples. In particular, the features in the measured data reflecting the effect of the interlayer coupling on $\Delta_1(\varphi)$ are correctly described by the developed theory [compare, e.g., Fig.~\ref{figure4}(g) and Fig.~\ref{figure1}(b)]. The fitting of the experimental data using Eq.~(\ref{eq:23}) allows to extract the value of the coupling constant $k$ (given in Table~\ref{table1}). The decrease in the spacer thickness from 9 to 4.5 nm strengthens $k$ from 1 to 180 Oe, which in turn enhances the in-plane anisotropy in $\Delta_1$: the difference between  the $\Delta_1$ values along and opposite to $\textbf{H}_{\textrm{b}}$ grows from essentially zero to 6 Oe ($\sim 6$\% of $\Delta_1$). An even more pronounced effect is observed in the sample with $d=3$ nm ($k \approx 690$ Oe, $\Delta_1^* \approx 23$ Oe), but we should note that the precision in determining the relevant multilayer properties in this strong-coupling case is not high [see the remark prior to Eq.~(\ref{eq:24})].

The use of Eq.~(\ref{eq:23}) makes it possible to estimate the damping parameters of F$_1$ and F$_2$, $\alpha_1$ and $\alpha_2$, respectively. For the case of $d=$ 6 nm, the obtained $\alpha_1$ and $\alpha_2$ values are $\sim 0.02$ and $\sim 0.05$, which are close to those reported in the literature for single Py and CoFe films ($\alpha_1 \sim 0.006 \div 0.02$ and $\alpha_2 \sim 0.05$).\cite{Tserkovnyak2002,Timopheev2014,Dubowik2002} The increase of the interlayer coupling affects the values of $\alpha_1$ and $\alpha_2$, but leaves the ratio $\alpha_2 / \alpha_1$ almost unchanged.

A closer look at $\Delta_1(\varphi)$ in Figs.~\ref{figure4}(b)--\ref{figure4}(d) reveals that there are peaks at $\varphi = \pm 90^{\circ}$ deviating from the predicted behavior (deviating significantly for $d=$ 6 nm). Although, a suitably detailed discussion of this finer structure goes beyond the scope of this paper, we would like to offer a suggestion as to the possible mechanism involved. Namely, anisotropic two-magnon scattering discussed in Subsection~\ref{C}. The source of this type of two-magnon scattering may be related to networks of inhomogeneities with some spatial orientation, formed as a result of the film deposition under a relatively strong external magnetic field needed to induce the exchange-pinning by the antiferromagnet. It was shown in Refs.~\onlinecite{Arias2000,McMichael2002} that for films with parallel steps or grooves, the two-magnon scattering mechanism makes the FMR linewidth strongly increase in the directions perpendicular to the step edges (grooves). Another mechanism that should be kept in mind in this regard is the acoustical and optical collective spin-resonance modes nominally expected in bi-layers with intermediate-strength coupling (vanishing for zero and strong coupling).\cite{Konovalenko2009} Such out-of-phase and in-phase mutual oscillations of the two ferromagnetic layers may cause additional dissipation for intermediate $k$ values, a detailed treatment of which requires a separate study.

\section{Conclusions}
	
The FMR linewidth and its anisotropy is studied experimentally and analyzed theoretically for F$_1$/f/F$_2$/AF multilayers, where spacer f has a low Curie point compared to the strongly ferromagnetic F$_1$ and F$_2$.

The role of the interlayer exchange coupling in the spin relaxation processes is investigated by varying the thickness of the spacer. It is shown that stronger interlayer coupling for thinner spacers enhances the microwave energy exchange between the outer ferromagnetic layers, with the magnetization of F$_2$ exchange-dragged by the resonant precession in F$_1$. A simultaneous but weaker inverse effect occurs: the magnetization of F$_1$ can be exchange-dragged by the precession in F$_2$, which leads to anti-damping and narrower FMR linewidths.

Strong interlayer coupling leads to strongly anisotropic magnetization damping, reaching its maximum for the direction antiparallel to the exchange-bias in the system.

By theoretically fitting the measured FMR data, the different contributions to the magnetic relaxation in the system are separated and discussed. Key physical parameters, such as the interlayer coupling constant and the in-plane anisotropy of the FMR linewidth, are quantified.

It is shown that in addition to the FMR relaxation effects related to the interlayer coupling, dispersion of the magnetic anisotropy fields in all of the layers can contribute to the FMR linewidth of F$_1$. Quantitative data for the dispersion parameters of the multilayer are obtained.

These results should be useful for designing high-speed nanodevices based on spin-thermionic control.

\begin{acknowledgments}
Support from the Swedish Stiftelse Olle Engkvist Byggm\"{a}stare, the Swedish Research Council (VR grant 2014-4548), the Science and Technology Center in Ukraine (project P646), and the National Academy of Sciences of Ukraine (projects 0115U003536 and 0115U00974) are gratefully acknowledged.
\end{acknowledgments}

\bibliography{manuscript_library}

\end{document}